\documentclass[10pt]{article}
\usepackage{latexsym,amsmath,amsfonts,graphicx,epsfig,url}

\def\be{\begin{equation}}
\def\ee{\end{equation}}
\def\bea{\begin{eqnarray}}
\def\eea{\end{eqnarray}}

\newcommand{\qed}{\rule{7pt}{7pt}}
\newcommand{\ba}{\begin{array}}
\newcommand{\ea}{\end{array}}

\newcommand{\ket}[1]{|#1\rangle}

\newcommand{\ignore}[1]{}

\begin{document}

\title{Fault-Tolerant Computing \\ With Biased-Noise Superconducting Qubits}

\author{P.~Aliferis\thanks{IBM Watson Research Center, Yorktown Heights, NY 10598.} \thanks{These authors contributed equally to this work.} , F.~Brito$^{*\dagger}$, D.~P.~DiVincenzo$^*$, J. Preskill\thanks{Institute for Quantum Information, California Institute of Technology, CA 91125.} , M.~Steffen$^*$, and B.~M.~Terhal$^*$}

\maketitle
\vspace{-.6cm}
\begin{abstract}
We present a universal scheme of pulsed operations for the IBM oscillator-stabilized flux qubit comprising the
controlled-$\sigma_{\rm z}$ ({\sc cphase}) gate, single-qubit preparations and measurements. Based on numerical simulations, we
argue that the error rates for these operations can be as low as about $.5\%$ and that noise is highly biased, with phase errors
being stronger than all other types of errors by a factor of nearly $10^{3}$. In contrast, the design of a controlled-$\sigma_{\rm x}$ ({\sc cnot}) gate for this system with an error rate of less than about $1.2\%$ seems extremely challenging. We propose a special
encoding which exploits the noise bias allowing us to implement a {\em logical} {\sc cnot} gate where phase errors and all other types of errors have nearly balanced rates of about $.4\%$.
%
% Removed:
% The principles underlying our scheme can also find application in other
% solid-state qubits, strengthening our hope that the fidelities
% needed for fault-tolerant quantum computation can be achieved with
% integrated circuits.
% Added:
Our results illustrate how the design of an encoding scheme can be adjusted and optimized according to the available physical operations and the particular noise characteristics of experimental devices.
% End
\end{abstract}
%\pacs{03.67.Hk, 03.67.-a, 03.67.Dd, 89.70.+c} \maketitle

%---------------------------------------------------------------------------%

\vskip .6cm After years of painstaking labor, superconducting qubits
\cite{Nori} are taking shape as viable elements for the construction
of a scalable quantum computer. Since the initial demonstrations of
coherent quantum dynamics in superconducting qubits
\cite{Nakamura,Martinis,Saclay,Chiorescu}, it has been recognized
that these systems have great potential versatility
\cite{Walraff,Steffen,Koch}, so that one can genuinely envision a
quantum-computing integrated circuit emerging from this research.
However, no clear way forward has been announced, owing largely to
one undeniable feature of large-scale quantum computation: it will
require a very high degree of fidelity in the execution of quantum
operations, much higher than has been reported in any present
experiments.

How high a fidelity, or how low an error rate, will be needed? On
the basis of fundamental early theoretical work
\cite{Aharonov96,Knill96b,Preskill97}, lip service is frequently
paid to a necessary universal set of operations containing the
two-qubit controlled-$\sigma_{\rm x}$ ({\sc cnot}) gate, and a
necessary ``threshold'' error rate in the $10^{-5}{-}10^{-4}$ range.
Some recent modeling for superconducting qubits \cite{Jamie,Fazio}
suggests that such noise levels could conceivably be reached in the
lab; however, in current experimental practice the ability even to
reliably {\em detect} such small error rates, let alone to achieve
them, is in fact very questionable.

In this paper, we will consider the possibility of constructing a
universal set of operations for the IBM ``Koch
qubit''\cite{Koch1,Koch} \footnote{After Roger Hilsen Koch
(1950-2007), the leader of the experimental superconducting-qubit
effort at IBM Research until his sudden death on August 4, 2007.}.
Although our set of operations will not contain the {\sc cnot} gate,
we will propose an encoding scheme for implementing {\em logical}
{\sc cnot} gates which can then be used for fault-tolerant quantum
computation. With the combination of our encoding scheme and other
improvements in the theory of quantum fault-tolerance
\cite{Knill05,Aliferis07b,Raussen07}, error rates for our elementary
operations in the $.1\%{-}.5\%$ range are expected to be tolerable
for practical quantum computation.
% Added:
We estimate using numerical simulations that error rates in this
range are possible for the Koch qubit. Though these estimates are
far from the error rates that have been currently measured in experiments, 
we hope our results will motivate and stimulate the research in superconducting
qubits in the IBM lab and elsewhere. More generally and even beyond superconducting qubits, our encoding scheme illustrates how techniques of quantum error correction and fault tolerance can be tailored to the available physical operations and the particular noise characteristics of experimental devices.
% End

Our proposal involves a synthesis of recent experimental and
theoretical developments. On the experimental side
\cite{Koch1,Koch}, pulsed operations for the Koch qubit were
discussed in \cite{Brito07}; the set of operations considered in
\cite{Brito07} included the controlled-$\sigma_{\rm z}$ ({\sc
cphase}) gate, the Hadamard ({\sc H}) gate, single-qubit
``diagonal'' rotations of the form $\exp(i\theta \sigma_{\rm z})$,
the preparation of a qubit in the state
$\ket{+}={1\over\sqrt{2}}(\ket{0}+\ket{1})$, and the measurement of
$\sigma_{\rm x}$ (where $\sigma_{\rm x}$, $\sigma_{\rm y}$, and
$\sigma_{\rm z}$ are the Pauli spin operators). For these
operations, \cite{Brito07} numerically estimated the error rates by
considering all physical sources of noise that we presently know how
to model for this system---$1/f$ flux noise, instrumental jitter in
pulse timing and amplitude, and Johnson noise from resistances in
the circuit. For the {\sc cphase} gate it estimated an error rate of
about $.45\%$, for the {\sc H} gate about $.4\%$, for diagonal
rotations about $.001\%$ for any $\theta$, for the preparation of
the state $\ket{+}$ about $.3\%$, and for the measurement of
$\sigma_{\rm x}$ about $.2\%$ \footnote{These numbers are obtained
by using the results in \cite{Brito07} and the definition of an
error rate in the Supplementary Material. In \cite{Brito07}, the
entanglement fidelity and not the error rate is used as a measure of
the noise strength.}.

\begin{figure}[t]
\begin{center}
\vspace{-1.0cm}
\includegraphics[ width=.75\columnwidth,keepaspectratio]{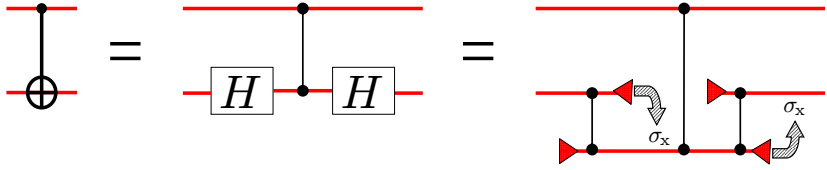}
\vspace{-6.8cm}
\end{center}
\caption{\label{fig:cnot} On the left, a {\sc cnot} gate is applied between two qubits. In the middle, the {\sc cnot} gate is simulated by using one {\sc cphase} and two {\sc H} gates. On the right, another simulation where the two {\sc H} gates are implemented by ``teleportation'' using ancilla qubits and measurements. Vertical lines with dots on both ends denote {\sc cphase} gates. A triangle pointing to the right denotes the preparation of a qubit in the state $|+\rangle$. A triangle pointing to the left denotes a measurement of $\sigma_{\rm x}$. Conditioned on the measurement outcomes, $\sigma_{\rm x}$ correction operators may be necessary as shown.
%This implies that $\sigma_{\rm z}$ errors preceding the measurements will be converted to $\sigma_{\rm x}$ errors on the outgoing qubits (which is a well-known property of the {\sc hadamard} gate).
         }
\end{figure}

In contrast, no direct implementation of a {\sc cnot} gate with error rate close to even $5\%$ could be devised for the Koch qubit. Alternatively, a {\sc cnot} gate can be implemented indirectly by composing other elementary operations. Fig.~\ref{fig:cnot} shows two possible methods; the first indirect implementation uses one {\sc cphase} and two {\sc H} gates, while the second one uses three {\sc cphase} gates, two qubits prepared in the state $\ket{+}$, and two measurements of $\sigma_{\rm x}$. Therefore, we estimate that the first indirect implementation would have an error rate of about $1.25\%$, and the second one about $2.3\%$.

Based on the modeling in \cite{Brito07}, it was further observed that noise in the {\sc cphase} gate acted predominantly as dephasing; {\em i.e.}, it affected the relative {\em phases} in the wavefunction giving rise to ``phase'' errors which can be expressed as diagonal matrices in the computational basis. All other errors, including errors due to relaxation and also ``leakage'' errors associated with transitions to states outside the computational space, were observed to be about an order of magnitude weaker. The same observation also applies to single-qubit diagonal rotations but, on the other hand, no similar bias was observed for the noise in the {\sc H} gate.

On the theoretical side, the findings of \cite{Brito07} motivated us
to revisit a longstanding problem in the theory of fault-tolerant
quantum computation: how to formulate an encoding scheme that
exploits large asymmetries in the {\em structure} of noise in
elementary operations. Here and in our companion paper
\cite{Aliferis07d}, we present such a scheme based on encoding the
noisy qubits using a repetition code, and we show how fault-tolerant
{\em logical} {\sc cnot} gates can be implemented for this code. The
intuition we gained from \cite{Brito07} is that highly biased noise,
with phase errors being much stronger than all other types of
errors, is possible for gates such as the {\sc cphase},  but it
should not be physically expected for generic operations.
Furthermore, an encoding scheme has to address the problem that a
noise bias can easily be lost as elementary operations are composed
together.

Our solution is to choose a universal set of elementary operations whose implementation induces noise that is biased towards dephasing, and where all gates commute with $\sigma_{\rm z}$ so that the noise bias is maintained. The operations we will use are the preparation of the state $\ket{+}$, the controlled-$\sigma_{\rm z}$ ({\sc cphase}) gate, and measurements of observables in the equator of the Bloch sphere, {\em i.e.}, of the form $\exp(i\theta\sigma_{\rm z})\sigma_{\rm x}$ for certain angles \footnote{Alternatively, we may restrict to measurements of $\sigma_{\rm x}$, if we add to our set of elementary operations the preparation of the states $|{+}i\rangle = \exp(i{\pi\over 4}\sigma_{\rm z})\ket{+}$ and $|T\rangle = \exp(i{\pi\over 8}\sigma_{\rm z})\ket{+}$; this is the universal set considered in \cite{Aliferis07d}. For the Koch qubit, it is more natural to move all rotations of the form $\exp(i\theta\sigma_{\rm z})$ before the measurements since, as we will discuss, corrective single-qubit diagonal rotations are necessary in the implementation of every {\sc cphase} gate. }. Noise for the preparation of the state $\ket{+}$ is naturally biased since $\ket{+}$ is an eigenstate of $\sigma_{\rm x}$. The structure of noise for the {\sc cphase} gate depends on the physical implementation and it can be engineered to be biased; here, we propose such a biased-noise implementation for the Koch qubit based on {\em adiabatic} pulses so that transitions between the computational-basis states are strongly suppressed. Finally, noise in a single-qubit measurement can be described by errors preceding the ideal measurement which need not have any specific structure since measurement has a classical output. %For all these operations, the remaining weak component of noise apart from dephasing is due to relaxation during the operations, and also due to leakage.

These theoretical ideas may have implications for the experiments
with the Koch qubit, and solid-state qubits in general. They suggest that a profitable
focus of experiments could be to design qubits with long relaxation
time $T_1$. Provided this is achieved, dephasing, which is the
dominant source of noise, can be much more effectively suppressed by
using our encoding scheme. Furthermore, the implementation of
low-noise {\sc cnot} or {\sc H} gates is not necessary; it suffices
to implement {\sc cphase} gates with highly biased noise, together
with single-qubit preparations and measurements. For the Koch qubit, eliminating the need to implement the {\sc H} gate
allows us to re-examine the pulse schemes in \cite{Brito07}, and to
find a simpler implementation of the {\sc cphase} gate for which
noise is much more biased than in \cite{Brito07}.
% Added:
Similar simplifications may be possible in other types of superconducting qubits. In particular, a standard experimental approach to suppressing dephasing noise is to intersperse quantum gates with additional corrective ``hardware'' operations such as spin echo pulses. Our encoding scheme can be seen as a complementary solution where the hardware operations are restricted to a much smaller set and dephasing is suppressed at the ``software'' level by using error correction applied on blocks of several noisy qubits.
% End

We will now discuss the details of our proposal. The central
question is how to devise a strategy for fault-tolerant quantum
computation that effectively exploits a bias in the noise of the
{\sc cphase} gate. To develop some intuition, let us assume for the
moment that there is no leakage, that independent phase errors occur
with probability $\varepsilon$, and that all other types of errors
occur with probability $\varepsilon' \ll \varepsilon$. Then a clear
strategy for the lowest-level coding of quantum data is to use an
$n$-qubit repetition code. One logical qubit is encoded in a
``block'' of $n$ physical qubits, a logical $\ket{+}$ state is
represented by all $n$ qubits being in the state $\ket{+}$, and a
logical $\ket{-}$ by all $n$ qubits being in the state $\ket{-}$.
Since the logical $\sigma_{\rm z}$ operator is a $\sigma_{\rm z}$
acting on all $n$ qubits in the block, phase errors on more than
half of the qubits in the block are necessary for a logical error to
occur after error correction; taking $n$ odd, and assuming that for
each qubit there are $t$ time steps where a phase error may occur,
the probability of a logical error is approximately
\begin{equation}
\varepsilon_{L}^{\;} \approx {n \choose {n+1 \over 2}} \left( t\, \varepsilon \right)^{n+1\over 2} \; .
\label{e1}
\end{equation}
On the other hand, the logical $\sigma_{\rm x}$ operator is a $\sigma_{\rm x}$ acting on {\em any} qubit in the block, so that even a single error different than a phase error on any qubit and at any time step cannot be corrected; thus, the probability of a logical error is approximately
\begin{equation}
\varepsilon'_L \approx n\, t\, \varepsilon' \; .
\label{e2}
\end{equation}
If the noise ``bias'' $\varepsilon/\varepsilon'$ is large, we can choose some large $n$ and obtain a significant reduction of the {\em logical} error rate; {\em e.g.}, if the bias is $10^3$ and even for $t\cdot \varepsilon = 5\%$, we find that $\varepsilon^{\,}_L\approx\varepsilon'_{L}<3.5\times10^{-4}$ by setting $n=7$. This sends an optimistic message, since several schemes are known for effectively implementing fault-tolerant quantum computation with error rates of order $10^{-4}$ \cite{Cross07}.

As we will discuss below, our pulsed implementation of a {\sc cphase} gate can in fact be optimized to lead to a noise bias of order $10^3$. But first, we must examine more closely whether there is actually a gain in using the repetition code as our naive argument so far indicates. The main concern is that a large intrinsic noise asymmetry in our elementary operations might be spoiled as these operations are composed together.

\begin{figure}[t]
\begin{center}
\begin{tabular}{c}
\vspace{-1cm}
\put(  0,275){(a)}
\put(270,295){(b)}
\includegraphics[ width=0.85\columnwidth,keepaspectratio]{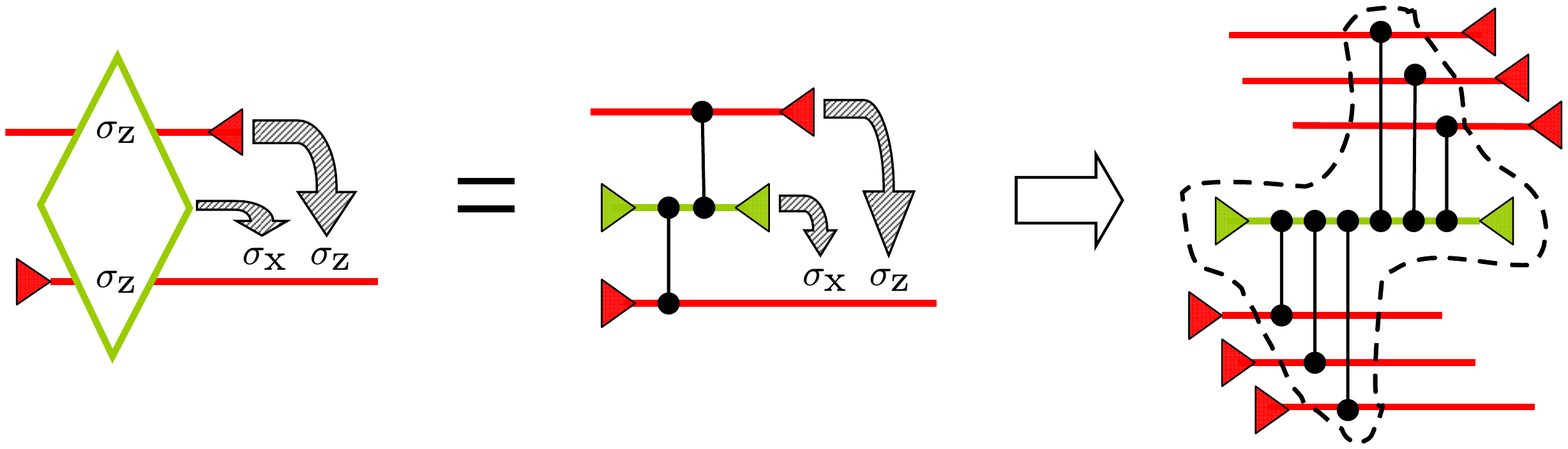}
 \vspace{-6.2cm}
\end{tabular}
\end{center}
\caption{\label{fig:tel} (a) On the left, an identity circuit which consists of preparing a qubit in the state $\ket{+}$, measuring $\sigma_{\rm z}\otimes \sigma_{\rm z}$ on this qubit and the input qubit, and finally measuring $\sigma_{\rm x}$ on the input qubit. After applying the correction operators shown conditioned on the measurement outcomes, the state of the input qubit is ``teleported'' to the output qubit. On the right, the measurement of $\sigma_{\rm z}\otimes \sigma_{\rm z}$ is implemented by using an ``ancilla'' qubit (shown in green) which interacts with the other two ``data'' qubits (shown in red) via {\sc cphase} gates before it is measured. (b) The circuit in (a) where the data qubits are encoded in the $n$-bit repetition code (here $n=3$); the ancilla qubit remains unencoded, but the subcircuit enclosed by the dashed curve must be repeated sequentially several times, and the majority of the measurement outcomes must be taken in order to correct errors (the repetitions are not shown). As in (a), logical $\sigma_{\rm z}$ or $\sigma_{\rm x}$ correction operators (not shown) may be necessary on the output data block.
         }
\end{figure}

For example, consider the problem of implementing a logical {\sc cnot} gate between two blocks of encoded qubits. The logical {\sc cnot} gate for the repetition code can be implemented by bitwise {\sc cnot} gates, and each {\sc cnot} gate can be simulated by using {\sc cphase} gates as in Fig.~\ref{fig:cnot}. But then due to the {\sc H} gates, phase errors that occur during the simulation of each {\sc cnot} gate are converted into errors of other types; {\em e.g.}, a $\sigma_{\rm z}$ error during the implementation of a {\sc H} gate will be converted to some linear combination of a $\sigma_{\rm z}$, a $\sigma_{\rm x}$, and a $\sigma_{\rm y}$ error. So, this construction of a logical {\sc cnot} gate destroys the asymmetric structure of the noise and nullifies the effectiveness of the repetition code. Avoiding this interconversion of noise is possible, but it requires a circuit of greater complexity. At the heart of this circuit construction is the identity operation in Fig.~\ref{fig:tel}(a); it consists of preparing a qubit in the state $\ket{+}$, measuring $\sigma_{\rm z}\otimes \sigma_{\rm z}$ on this qubit and the input qubit, and finally measuring $\sigma_{\rm x}$ on the input qubit.

The basic idea is now to encode the qubits in the repetition code. We have already mentioned that the preparation of a logical $\ket{+}$ involves simply preparing every qubit in the block in the state $\ket{+}$. A measurement of the logical $\sigma_{\rm x}$ is also very simple and robust; it can be implemented by measuring $\sigma_{\rm x}$ on each qubit in the block, and then taking the majority of the outcomes to correct errors. It remains to discuss how to implement a measurement of the logical $\sigma_{\rm z}\otimes \sigma_{\rm z}$ on the two blocks. The difficulty is that the measurement must be robust against the dominant phase errors, and it must also respect the biased structure of the noise. Fig.~\ref{fig:tel}(b) shows our solution: A single unencoded ``ancilla'' qubit is prepared in the state $\ket{+}$, and then {\sc cphase} gates are applied between this ancilla qubit and all other qubits in the two blocks. Finally, a measurement of $\sigma_{\rm x}$ is performed on the ancilla, and a logical $\sigma_{\rm x}$ correction may be necessary on the output block conditioned on the measurement outcome. Since even a single phase error on the ancilla can cause an error in the measurement outcome, the measurement of the logical $\sigma_{\rm z}\otimes \sigma_{\rm z}$ must be repeated sequentially several times; by taking the majority of the outcomes we can suppress the probability of a logical error at the output.

\begin{figure}[t]
\begin{center} \hspace{3cm}
\includegraphics[ width=0.58\columnwidth,keepaspectratio, angle=-90]{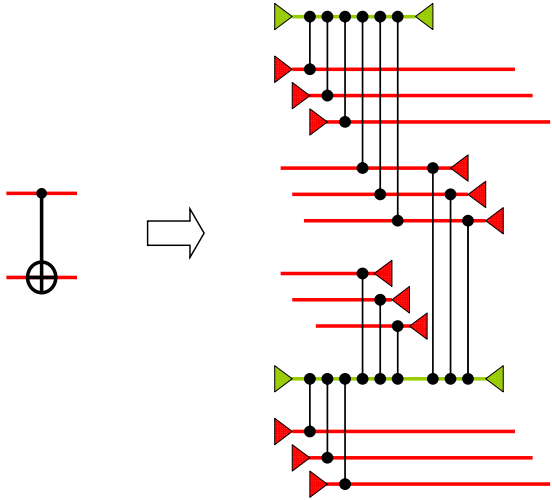}
\vspace{-2.7cm}
\end{center}
\caption{\label{fig:enc_cnot} A logical {\sc cnot} gate between two blocks encoded in the repetition code (here again, the 3-bit code). Ancilla qubits (shown in green) are prepared in the state $|+\rangle$, they interact via {\sc cphase} gates with the data qubits (shown in red), and they are measured along the eigenbasis of $\sigma_{\rm x}$. As in Fig.~\ref{fig:tel}, the measurements with ancilla qubits must be repeated sequentially several times so that errors can be corrected by taking the majority of the outcomes (the repetitions are not shown). Finally, $\sigma_{\rm x}$ is measured on each qubit in the two input blocks, and the majority is taken on each block. Conditioned on the results of the majorities, logical correction operators (not shown) may be necessary on the output blocks \cite{Aliferis07d}.
         }
\end{figure}

Using the building blocks just discussed, it is possible to construct a fault-tolerant logical {\sc cnot} gate that respects the physical-level bias of the noise; see Fig.~\ref{fig:enc_cnot}. This circuit implements a logical {\sc cnot} gate between the two blocks encoded in the repetition code, while at the same time the logical state of each block is teleported to a new block and phase errors are corrected. Because of the use of teleportation, this circuit has the additional feature that it largely prevents the propagation of leakage errors. In Fig.~\ref{fig:enc_cnot}, the output data qubits always interact with an ancilla qubit prior to any interaction with the input data qubits; therefore, there is no possibility for leakage to propagate from the input to the output qubits. And furthermore, if we consider implementing a sequence of logical {\sc cnot} gates, we observe that every qubit is eventually measured after only a small, fixed number of time steps, and the measurement effectively converts leakage to regular qubit errors \cite{Aliferis05c}.

In addition, it is possible to construct with the same building blocks a fault-tolerant preparation of the logical $\ket{0}$ and $\ket{+}$ states, and a measurement of the logical $\sigma_{\rm z}$ and $\sigma_{\rm x}$ for the repetition code. The logical {\sc cnot} gate, the preparation of logical $\ket{0}$ and $\ket{+}$, the measurement of logical $\sigma_{\rm z}$ and $\sigma_{\rm x}$, and single-qubit measurements of $\exp(i\theta\sigma_{\rm z})\sigma_{\rm x}$ for $\theta=\pi/4$ and $\theta=\pi/8$ suffice for implementing universal fault-tolerant quantum computation; the full scheme is discussed in our companion paper \cite{Aliferis07d}.

\begin{figure}[t]
\begin{center}
\includegraphics[ width=0.65\columnwidth,
 keepaspectratio]{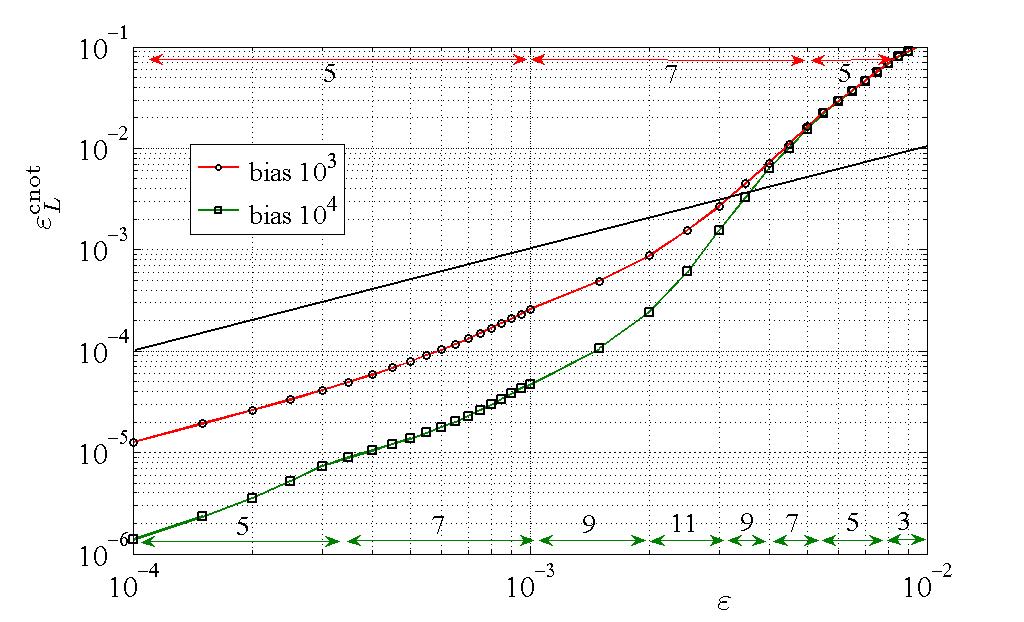}
\vspace{-.7cm}
\end{center}
\caption{\label{fig:plot} Upper bounds on the total probability $\varepsilon^{\rm cnot}_{L}$ of logical errors for the logical {\sc cnot} gate as a function of the probability $\varepsilon$ of physical-level phase errors for a bias of $10^3$ and $10^4$. %; our upper bounds for the probability of all other types of errors for the logical {\sc cnot} gate are equal to the upper bound for the probability of logical phase errors, so we expect that the bias is largely eliminated at the logical level.
To obtain our bounds, we have optimized over the block size $n$ of the repetition code, and the number $k$ of repetitions of measurements with ancilla qubits; the optimal choice is to have $n=k$, while the optimal value depends on $\varepsilon$ and the bias as shown.
The straight line with slope unity serves as a guide to the eye.
         }
\end{figure}

Since our building blocks only use {\sc cphase} gates, $\ket{+}$ preparations, and $\sigma_{\rm x}$ measurements, and since phase errors commute with {\sc cphase} gates, it is simple to estimate the error rates for the logical {\sc cnot} gate given the physical-level error rates. In \cite{Aliferis07d}, we give upper bounds for the logical error rates in closed form as a function of the block size $n$ of the repetition code and the number of repetitions $k$ of the measurements executed with ancilla qubits. The outcome is qualitatively as in Eqs.~(\ref{e1}) and (\ref{e2}) with $t=c\,k$ for some constant $c\approx 2$ or 3. It follows from this analysis that if the noise bias in our elementary operations is about $10^{3}$ or greater, encoding in the repetition code is effective and logical errors are significantly weaker than the physical-level phase errors for $\varepsilon$ of order $.1\%$ or smaller; see Fig.~\ref{fig:plot} \footnote{Ref.~\cite{Aliferis07d} also discusses a more sophisticated procedure for decoding the repetition code which is effective for $\varepsilon$ of about $.5\%$ or smaller. For simplicity, we do not analyze this decoding procedure in this paper.}.

\begin{figure}[t!]
\begin{center}\includegraphics[ width=1\columnwidth,keepaspectratio]{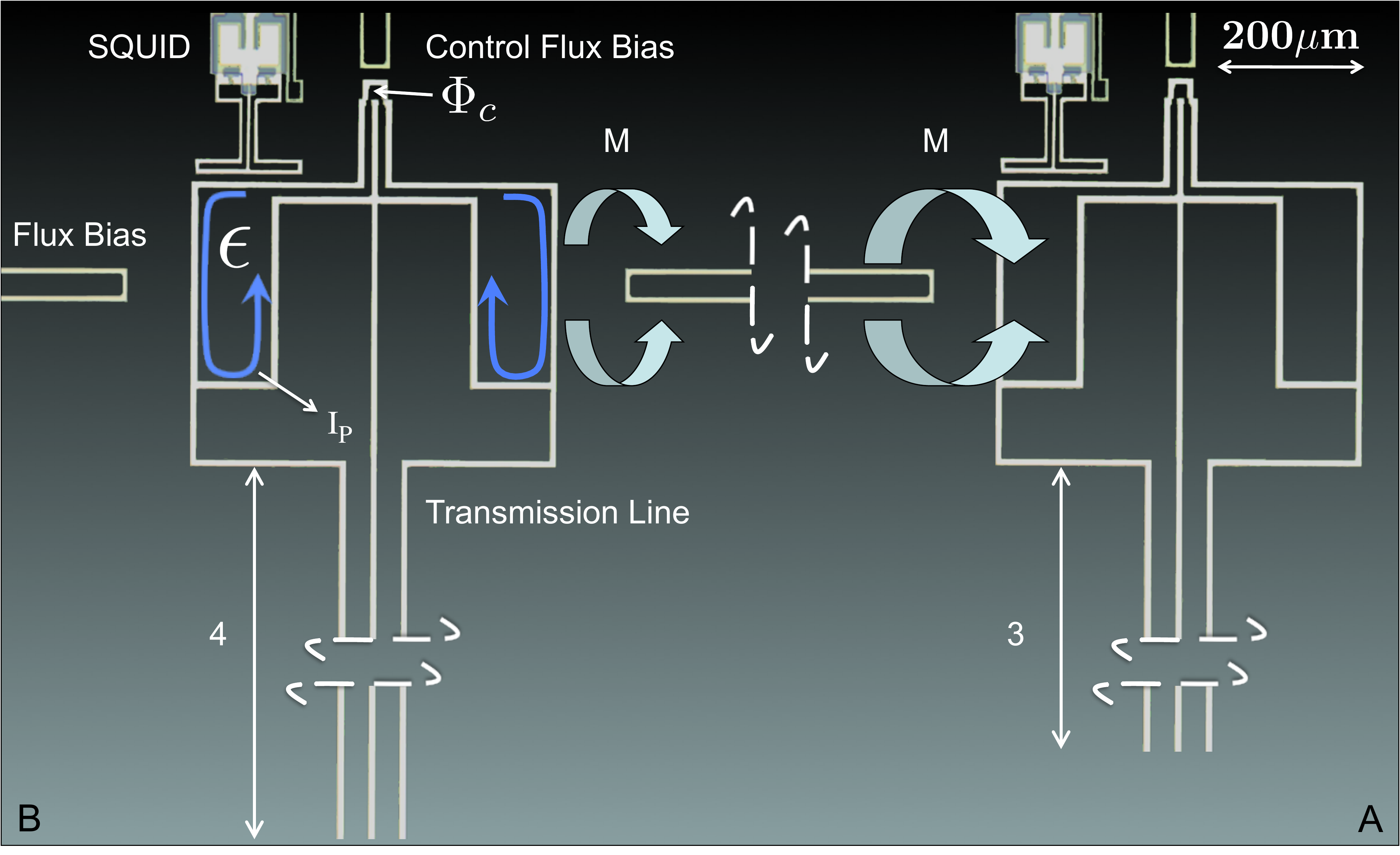}
\vspace{-.8cm}
\end{center}
    \caption{\label{fig:1} Physical layout of two Koch qubits \protect\cite{Koch} coupled as needed to implement a {\sc
cphase} gate. Qubits A and B differ only in the resonant frequencies of the transmission line resonators attached to them. These frequencies are controlled by the physical length of the transmission lines, and are in the ratio 3:4.  The single-qubit device Hamiltonians can be varied independently by controlling the ``control flux'' $\Phi_C$ corresponding to the flux through the small loop, and the ``flux bias'' $\epsilon$ corresponding to the flux difference in the two large loops. The SQUIDs perform quantum measurements on the states of the individual qubits.  The qubits are coupled via a fixed mutual inductance $M$ to a superconducting loop. For small values of $\Phi_C$, the basis states $\ket{0}$ and $\ket{1}$ of each qubit correspond to different orientations, counterclockwise and clockwise, %respectively,
of the persistent current $I_P$.
         }
\end{figure}

\begin{figure}[t]
\begin{center}\includegraphics[ width=0.55\columnwidth,keepaspectratio]{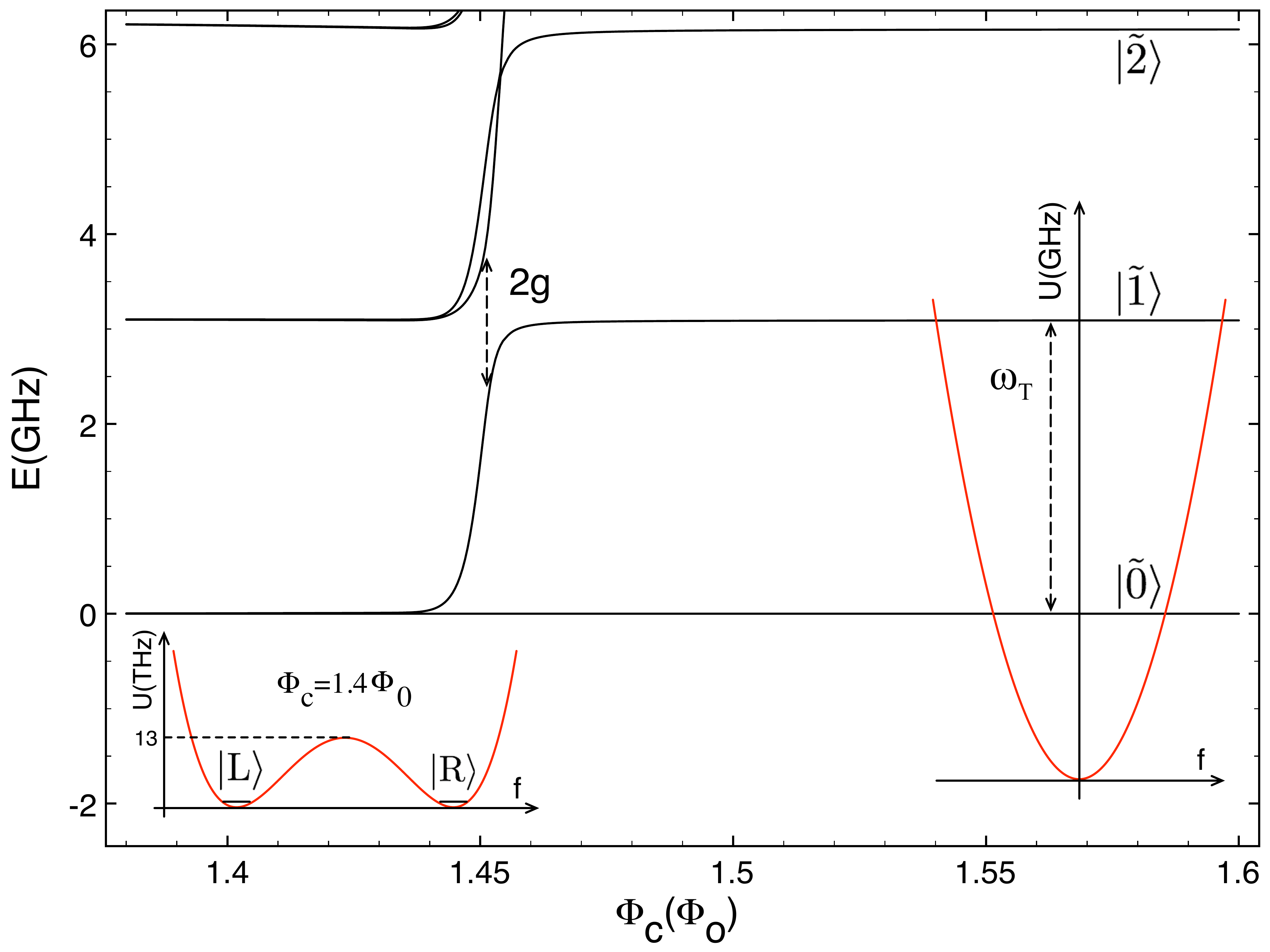}
\vspace{-.5cm}
\end{center}
\caption{\label{fig:2} The lowest energy levels of a single Koch qubit on the ``symmetric line'' corresponding to $\epsilon=0$, as a function of $\Phi_C$.  For small $\Phi_C$, $\Phi_C=1.4\Phi_0$ ($\Phi_0=h/2e$), the ground state is doubly degenerate; the degeneracy may be lifted by setting $\epsilon\not = 0$, allowing for qubit measurement and preparation. The two basis states $|0\rangle$ and $|1\rangle$ correspond to the two states $\ket{L}$ and $\ket{R}$ of a double-well potential with a very large potential barrier between them as sketched, which describe the two different orientations of the persistent current $I_P$ in the flux qubit; the rescaled dynamical variable $f$ is explained in \protect\cite{Brito06}. For both $|0\rangle$ and $|1\rangle$, the superconducting transmission line is in its ``vacuum'' 0-photon state. As $\Phi_C$ is increased to around $1.45\Phi_0$, the barrier height drops, leading to a rapid increase in the tunnel splitting between the two lowest states. When the tunnel splitting equals the transmission-line resonator frequency $\omega_T$, there is an avoided level crossing with splitting $2g$. For larger $\Phi_C$, the qubit is ``parked"; now, the two lowest energy states $\ket{\tilde 0}$ and $\ket{\tilde 1}$ correspond to the 0- and 1-photon states of the transmission line respectively, their energies are independent of $\Phi_C$, and their effective potential is highly harmonic as sketched. In this regime, for both $\ket{\tilde 0}$ and $\ket{\tilde 1}$, the flux-qubit state is the symmetric state $\ket{S}\equiv {1\over \sqrt{2}} (\ket{L} + \ket{R})$.
         }
\end{figure}

The noise bias reported in \cite{Brito07} was about a factor of 10, which has motivated us to re-examine whether greater levels of bias are conceivable for the Koch qubit. As we will now discuss, the noise bias can, in fact, be dramatically improved at the expense of a very minor increase in the rate of phase errors. Fig.~\ref{fig:1} shows the structure of the Koch qubit \cite{Koch1,Koch}. It is nominally a ``flux'' qubit, meaning that the computational basis states $\ket{0}$ and $\ket{1}$ are quantum states corresponding to distinct circulating-current orientations; see Fig.~\ref{fig:1}. The mode of operation is highly tunable via an external ``control flux'' $\Phi_c$ threading the small loop; see Fig.~\ref{fig:2}. At small $\Phi_c$, the degeneracy of the two circulating-current states is lifted by the ``flux bias'' $\epsilon$ corresponding to the flux difference in the two large loops, and the two basis states are easily detected and initialized but not very phase-coherent. As $\Phi_c$ is increased, the barrier decreases until it eventually disappears, and tunnel splitting between the basis states turns on rapidly. At even larger $\Phi_c$, we enter an almost harmonic single-well regime, but before this another essential element of the Koch qubit enters the picture.

The flux-qubit states are strongly coupled to the fundamental mode of a superconducting transmission-line resonator, so that $\ket{0}$ and $\ket{1}$ pass via an avoided level crossing to $\ket{\tilde 0}$ and $\ket{\tilde 1}$ corresponding to the 0- and 1-photon modes of the oscillator. The superconducting transmission line is highly phase-coherent \cite{Koch}. In the IBM experiments we have seen {\em circa} 50,000 Ramsey fringes associated with these states corresponding to $T_2 = 2.5\mu{\rm sec}$ \cite{IBMunpub}; we expect much longer $T_2$ times to be possible. This fact has led us in \cite{Brito07} to the following strategy for implementing low-noise operations. When not being acted upon, quantum information is stored in the highly-coherent oscillator energy levels; we then say that the qubit is ``parked.'' All needed operations are done by {\em adiabatic} pulsing out of parking. Each flux qubit has a fixed, untuneable two-qubit coupling to a set of nearby flux qubits in order to implement two-qubit gates. To assure that the effective coupling between parked qubits is negligible, these couplings are to be only between qubits with {\em different} resonator frequencies, so that resonant transfer of photons between different resonators does not take place in the parked state. Below we examine the simplest scheme with just two resonator frequencies, $3.1$GHz and ${3 \over 4}\cdot 3.1$GHz (the commensurability of these frequencies aids in the maintenance of rotating frames for these qubits). Correspondingly, we have two species of qubits, A and B respectively, and two-qubit couplings exist only between qubits of different species---this is not a severe restriction since, in our encoding scheme, the only interactions are between data qubits (red qubits in the figures) and ancilla qubits (green qubits).

\begin{figure}[t]
\begin{center}
\includegraphics[ width=.7\columnwidth,keepaspectratio]{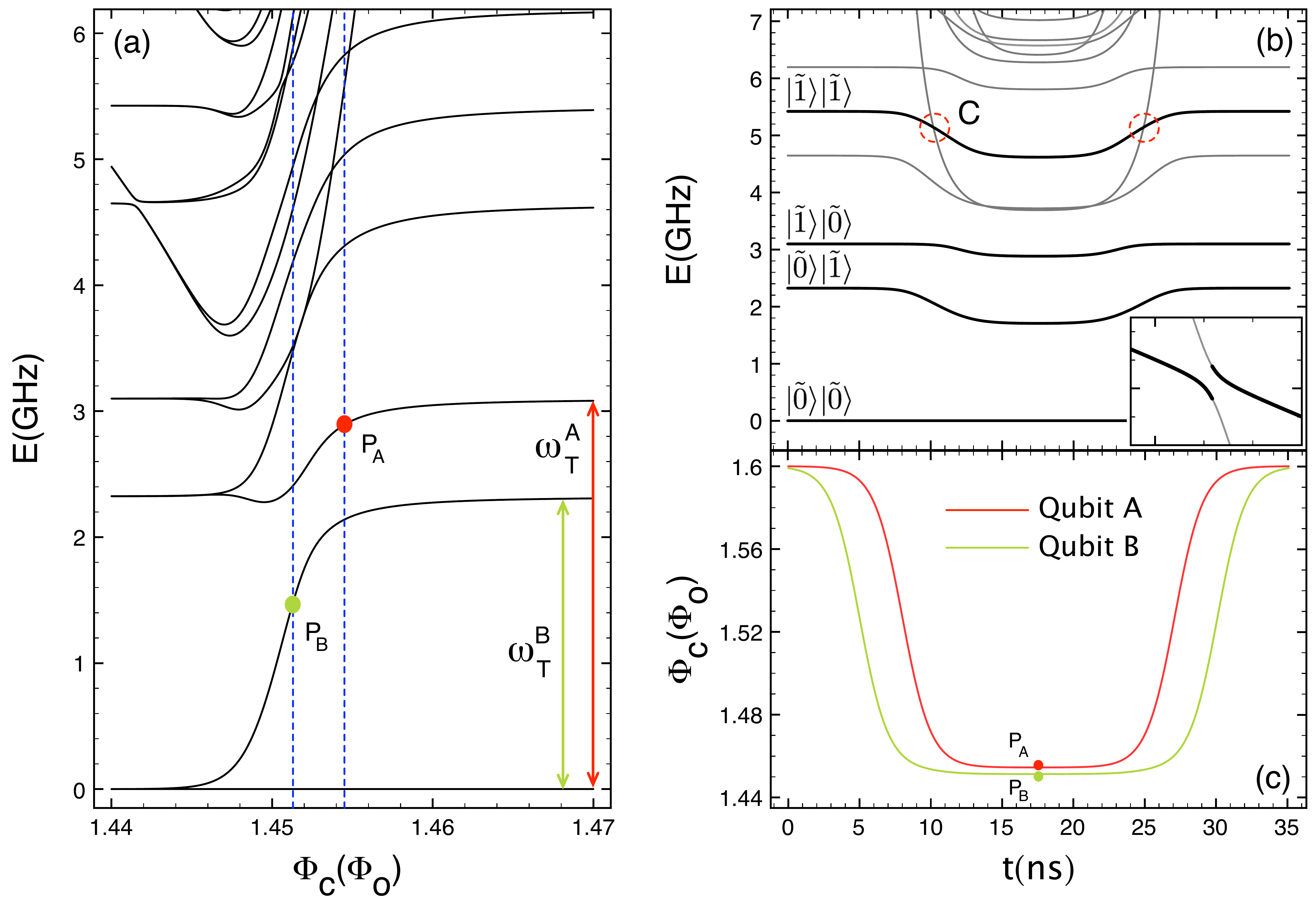}
\vspace{-.5cm}
\end{center}
 \caption{\label{fig:band_flux} {\sc cphase} gate modeling.  (a) The energy levels of two coupled qubits for the special case $\Phi_C^A=\Phi_C^B$. (b) Variations of the eigenlevels of the two-qubit system for the optimal implementation of a {\sc cphase} gate. To very good approximation, the quantum evolution adiabatically follows the four states $\ket{\tilde 0}\ket{\tilde 0}$, $\ket{\tilde 0}\ket{\tilde 1}$, $\ket{\tilde 1}\ket{\tilde 0}$ and $\ket{\tilde 1}\ket{\tilde 1}$. A crucial moment in the evolution is at the crossing $C$ (and its time-reversed image in the second half of the pulse), when $\ket{\tilde 1}\ket{\tilde 1}$ crosses a state outside the computational space. %, leading the possibility of leakage.
Leakage to this state is suppressed because the avoided-crossing gap is very small (the inset shows the dispersion at $C$ magnified by 200,000x), and because the angle of crossing is made larger by delaying the onset of the $A$ pulse relative to the $B$ pulse. Apart from this state and the four computational-basis states, all other eigenstates can be entirely left out of our numerical modeling because of their large distance from the $\ket{\tilde 1}\ket{\tilde 1}$ and $\ket{\tilde 1}\ket{\tilde 0}$ states. (c) The optimal control-flux pulses for $A$ and $B$. Pulses are constructed as sums of tanh functions in order to have a standard smooth shape. It is found preferable to unpark $A$ much less deeply (to $P_a$) than $B$ (to $P_b$). Although $P_a$ and $P_b$ in (a) do not exactly correspond to the points $P_a$ and $P_b$ in (c) because $\Phi_C^A$ and $\Phi_C^B$ are not equal throughout the optimal pulses, they provide a good illustration of the relatively large energy difference ($\sim 1.5$GHz) corresponding to the small change in $\Phi_C$ ($\sim .004 \Phi_0$) between $P_a$ and $P_b$ in (c).
         }
\end{figure}

Fig.~\ref{fig:band_flux}(a) shows the energy levels of the coupled two-qubit system as a function of the control flux---assumed equal for the two qubits in the figure---with the flux bias held on the ``symmetric line'' $\epsilon = 0$ on which the effective qubit potential $U$ has a reflection symmetry as in Fig.~\ref{fig:2}. In the ``parking'' region $\Phi_c \geq 1.46 \Phi_0$, the effective interaction between the two qubits is strongly suppressed. The energy levels are essentially those of the unperturbed transmission line resonators and, because to very high accuracy we have energy additivity $E(\ket{\tilde 1}\ket{\tilde 1})=E(\ket{\tilde 0}\ket{\tilde 1})+E(\ket{\tilde 1}\ket{\tilde 0})$, there is no conditional phase accumulation in the two-qubit state. Out of parking, this energy additivity is violated and a {\em conditional} phase shift can accumulate. If only one qubit is pulsed out of parking, single-qubit ``diagonal'' rotations of the form $\exp(-i \theta \sigma_{\rm z})$ can be effected (in the frame of reference rotating at the resonator frequency). %; if these rotations precede a measurement of $\sigma_{\rm x}$, they allow for a measurement in any angle in the equator of the Bloch sphere to be implemented.
If two coupled qubits are pulsed out of parking, a {\sc cphase} gate can be implemented by choosing the pulse timing appropriately; at the same time, {\em known} single-qubit phase shifts accumulate which can be compensated by corrective diagonal rotations on each qubit \cite{Brito07}. %, but since these rotations commute with the {\sc cphase} gates, they can be model before the final measurements of each qubit to re-define the measurement angles $\theta$.

There is, however, one point of concern which reveals itself more clearly when we plot the energy levels of the same two-qubit system as a function of time during the implementation of a {\sc cphase} gate; see Fig.\ref{fig:band_flux}(b). At the energy-level crossing marked C, the avoided-crossing gap between the state $\ket{\tilde 1}\ket{\tilde 1}$ and a state {\em outside} the computational space becomes small, and this may potentially lead
to increased leakage errors. One possible strategy to avoid this problem would be to increase the avoided-crossing gap by changing the flux bias to $\epsilon\not = 0$ in order to move away from the symmetric line; this solution was adopted in \cite{Brito07}. However, departing from the symmetric line has its disadvantages. The lowering of symmetry makes the system less protected from low-frequency noise, and it also causes the effective relaxation time $T_1$ to be shortened making it harder to achieve a high bias in the noise.

It would therefore be desirable to perform the operation at the symmetric line. Our new observation is that the avoided-crossing gap is actually very small; it is never larger than 100kHz even taking parameter shifts
due to low-frequency noise into account. So for our pulse profiles, the Landau-Zener tunneling probability is extremely
close to one, meaning that we can pulse through the avoided-crossing gap and suffer essentially no leakage. After
trying many pulse designs, and checking the noise bias they produce, we find the pair of pulses in Fig.~\ref{fig:band_flux}(c) for the two species of qubits to work the best. We observe that in this pulsing scheme the low-frequency qubit B is unparked far more deeply (point ${\rm P}_{\rm b}$) than the high-frequency qubit A (point ${\rm P}_{\rm a}$); this choice is optimal because, if the two species of qubits were unparked symmetrically, the same level of dephasing noise could only be maintained by making the pulse duration significantly longer increasing the noise due to relaxation and leakage.

To estimate the effect of noise in the implementation of our {\sc cphase} gate, we have performed multiple computer simulations of the evolution of the two coupled qubits during the gate where, in each simulation, low-frequency noise is added to the ideal dynamics. To model low-frequency noise, in each simulation we take the actual applied flux and the actual timing in a pulse to deviate from the ideal by an amount which is chosen from a Gaussian distribution with
zero mean and variance $6\mu\Phi_0$ and $6p{\rm sec}$ respectively \cite{Brito07}. From each simulation we extract a unitary which describes the noisy implementation of the {\sc cphase} gate; by averaging over a large number of simulations which corresponds to integrating over the Gaussian fluctuations, we obtain a superoperator describing the noisy implementation of the gate. By expressing this superoperator as $\mathcal{N} \circ \mathcal{CZ}$ where $\mathcal{CZ}$ is the ideal superoperator when there is no noise, we can define an ``error rate'' in terms of the norm of $\mathcal{E} = \mathcal{N} - \mathcal{\hat I}$ where $\mathcal{\hat I}$ is a trace-decreasing superoperator proportional to the identity superoperator. Although this rate does not necessarily correspond to the probability of an error, it is possible to formulate a fault-tolerance analysis similar to the case of stochastic noise. In particular, we may expand $\mathcal{E} = \mathcal{E}_{\rm phase} + \mathcal{E}_{\rm other}$ where $\mathcal{E}_{\rm phase}$ contains the terms of $\mathcal{E}$ which are diagonal in the computational basis, and $\mathcal{E}_{\rm other}$ contains all other terms. Then, Eqs.~(\ref{e1}) and (\ref{e2}) remain unchanged if $\varepsilon$ and $\varepsilon'$ are re-interpreted as the norms of $\mathcal{E}_{\rm phase}$ and $\mathcal{E}_{\rm other}$ respectively. In the Supplementary Material, we give more details about this definition and about the derivation of the error rates that we discuss below.

Based on our modeling, we find the following hierarchy of expected
error levels for the {\sc cphase} gate: The rate for phase errors is
$1.96\times 10^{-3}$ for qubit A and $4.6\times 10^{-3}$ for qubit
B. For all other types of errors, the dominant contribution is due
to {\em relaxation} to the ground state during the $35.1n{\rm sec}$
duration of the {\sc cphase} gate; using our previous results based
on the Caldeira-Leggett model which give an estimate of $T_1\approx
10m{\rm sec}$ \cite{Brito06,Brito07}, we expect their rate to be
around $3.5\times 10^{-6}$ for both qubits A and B. Therefore, as
desired, there is a large contrast between the rates for phase
errors and all other types of errors. We must also finally assess
the magnitude of leakage. Since leakage errors cannot in general be
corrected by our repetition code, it is crucial that leakage is
suppressed to a very large extent; in fact, we have found that the
rate for leakage errors is approximately $3.5\times 10^{-6}$.

 %for qubit A is $1.6\times 10^{-8}$, for qubit B it is $3.5\times 10^{-6}$, while the rate for the simultaneous occurrence of leakage on both qubits is significantly smaller and we have neglected it. In addition, leakage propagation through {\sc cphase} gates is strongly suppressed; if one of the qubits prior to a {\sc cphase} gate is outside the computational space, the rate for leakage errors on the {\em other} qubit after the implementation of the gate is $3 \times 10^{-3}$.

We have also designed pulses for the preparation of a qubit in the
state $|+\rangle$, for single-qubit diagonal rotations, and for the
measurement of $\sigma_{\rm x}$; these pulses are only slightly
modified from \cite{Brito07}. Since the diagonal rotations commute
with the {\sc cphase} gate, they can always be moved to occur
immediately before the measurements of $\sigma_{\rm x}$; the
combined operation is a measurement of an operator of the form
$\exp(i \theta \sigma_{\rm z})\sigma_{\rm x}$. The rate for phase
errors in a preparation of $\ket{+}$ is $2.75\times 10^{-3}$ for
both qubits A and B, and the rate for all other types of errors is
$3.5\times 10^{-7}$. However, because a preparation of $|+\rangle$
is performed by a {\em non-adiabatic} pulse, leakage is
significantly larger than in the {\sc cphase} gate; the leakage
error rate is $3.77\times 10^{-7}$ for qubit A and $1.5\times
10^{-5}$ for qubit B. For a measurement of $\exp(i \theta
\sigma_{\rm z})\sigma_{\rm x}$, we can describe noise in terms of
effective errors of no specific structure preceding the ideal
measurement, and their rate is $1.83\times 10^{-3}$ essentially
independent of $\theta$.

Given these physical-level error rates, we can obtain {\em upper
bounds} on the error rates for the logical {\sc cnot} gate by using
the equations in \cite{Aliferis07d}. In this analysis, we use the
error rates we have computed by averaging over a large number of
simulations as fixed, constant error rates, and we will ignore any
correlations between the fluctuations around these average values in
space and time. The analysis must also consider leakage errors which
are not discussed in \cite{Aliferis07d}. Our method for analyzing
leakage is based on the following observations. First, as we have
already noted, the fact that teleportation is used to implement the
logical {\sc cnot} gate prevents leakage errors from propagating
across multiple logical gates. In the worst case, one leakage error
can propagate from the logical gate where it occurs to the following
one. But this can be prevented by inserting a logical teleportation
preceding every logical gate; {\em i.e.}, the logical state of each
output block from a logical gate is teleported to a new block as in
Fig.~\ref{fig:tel} before the next logical gate is applied. Then, a
leakage error that occurs in a logical gate or in the logical
teleportations preceding it can only affect this logical gate and no
other.

In our analysis of the logical {\sc cnot} gate, we have optimized
over the block size $n$ of the repetition code, and the number $k$
of repetitions of measurements with ancilla qubits. For our
physical-level error rates, the optimal choice is $(n,k)=(5,7)$,
where $k$ is larger than $n$ because qubits of species B (which are
chosen as the ancilla qubits) are more noisy that qubits of species
A (which are chosen as the data qubits). With this choice, we find
that the logical {\sc cnot} gate has nearly balanced rates for phase
errors and all other types of errors of at most $4.62\times 10^{-3}$
and $3.98\times 10^{-3}$ respectively.
%; more discussion about the derivation of these upper bounds is given in \cite{suppl}.

This is an improvement by a factor of about 3 over the best
alternative method we have for implementing a {\sc cnot} gate
without the encoding in the repetition code; as we have already
discussed, we have found no direct implementation of a {\sc cnot}
gate for the Koch qubit with an error rate better than $5\%$, and
simulating the {\sc cnot} gate indirectly as in Fig.~\ref{fig:cnot}
leads to balanced rates for phase errors and all other types of
errors of about $1.25\%$. Since our error-rate upper bounds also
apply to the other logical operations for the repetition code needed
for universality \cite{Aliferis07d}, our analysis indicates that our
estimated physical-level error rates for the Koch qubit are
tantalizingly close to those needed for effective fault-tolerant
operation; in the literature, the highest estimated error thresholds
are of order $1\%$, and the best proven thresholds are of order
$.1\%$ \cite{Knill05, Aliferis07b, Raussen07, Cross07}.

To conclude, 
% Removed:
% we have made the case that elementary operations for
% flux qubits can be expected to achieve error rates which allow for
% fault-tolerant quantum computation. 
% Added:
we have discussed an encoding scheme which has been especially tailored to the physical operations that are naturally available in the IBM qubit ({\sc cphase} gates, qubit preparations and measurements), and the noise characteristics that can be expected for this system according to our theoretical modeling. 
% End
We believe that
the basic principles underlying our proposal---primarily, the
implementation of {\em adiabatic} {\sc cphase} gates with highly
biased noise, and the suppression of phase errors by encoding in the
repetition code---can be successfully adapted to apply to other
promising systems besides flux qubits such as, {\em e.g.},
superconducting phase qubits \cite{Martinis,Steffen}.

Our estimated error rates incorporate contributions from all sources
of noise which are understood in experiments at present. However,
the experimental reality today is that noise is dominated by $T_1$
processes which are not fully understood, and coherence times are
significantly below the values we have obtained from our calculation
\cite{Brito07}. Our results should therefore be seen as preliminary
and suggestive. Certainly, simplifications in modeling low-frequency
$1/f$ noise have been made, so that noise fluctuations have been
assumed to be constant during the execution of each gate, and also
noise correlations and flux drifts across multiple gates have been
ignored; see \cite{Brito07}.
% Removed:
% Despite these simplifications, we believe that our model describes
% to good accuracy some of the essential features of noise in these
% devices.
% End

Furthermore, we should emphasize that there are several
possibilities for complementing the methods of ``software'' error
correction we have described here with ``hardware'' error correction
which is done directly at the physical level; {\em e.g.}, systematic
noise correlations and flux drifts over long time scales could be
suppressed by periodically recalibrating qubits off-line before they
are re-used, or superconducting qubits could be designed to have
physical error-correction properties as in \cite{Kitaev} which uses
the same set of elementary operations as ours. Finally, we should
note that the topology of interactions for our repetition-code
scheme is not attainable with short-range interactions on a square
lattice. We expect however that a greater but limited range of
interactions where each qubit is coupled to 10 or 20 other nearby
qubits would suffice. Various possibilities exist for implementing
such interactions for superconducting-qubit layouts where
crossovers, multiple couplers, and indirect couplings via
transmission lines are all available.

% Removed:
% More generally and even beyond superconducting qubits, our message
% is that future experiments could focus on improving the relaxation
% time $T_1$. Provided $T_1$ is long enough, dephasing noise can be
% suppressed by using the encoding scheme and fault-tolerant circuits
% we have described in this paper.
% End

DDV and BMT have been partly supported by IARPA under ARO Contract No. W911NF-04-C-0098.  JP is supported in part by DoE under Grant No. DE-FG03-92-ER40701, NSF under Grant No. PHY-0456720, and NSA under ARO Contract No. W911NF-05-1-0294.

%-----------------------------------------------------------------------------------------------------------%
\bibliographystyle{unsrt}

%------------------------------------------------------------------%
\appendix

\cleardoublepage

%------------------------------------------------------------------%
\begin{center}
{\LARGE Supplementary Material on Error Rates}

\vspace{.5cm}

{\large P.~Aliferis, F.~Brito, D.~P.~DiVincenzo, J. Preskill, M.~Steffen, and B.~M.~Terhal}
\end{center}

\vspace{1cm}
In our analysis, we model each noisy elementary operation by a superoperator. Therefore, we ignore temporal or spatial correlations between different operations, such as the correlations in time which are inevitably present for $1/f$ noise. We consider two sources of noise. First, $1/f$ noise in the control parameters---the applied fluxes, and the pulse synchronization---during the execution of each operation; $1/f$ noise leads primarily to dephasing and it will be modeled by a superoperator ${\cal N}_{1/f}$. Secondly, thermal relaxation noise which is continuously present; relaxation is the primary source of errors different than phase errors and it will be modeled by an amplitude-damping superoperator ${\cal N}_{T_1}$.

Leakage errors depend both on the implementation of our operations and also on $1/f$ noise. We recall that during qubit preparation and measurement, which are implemented by {\em non-adiabatic} pulses, leakage can arise from Landau-Zener transitions to excited levels in the flux-qubit potential; during the execution of a {\sc cphase} gate, leakage primarily occurs at the energy-level crossing marked C in Fig.~\ref{fig:band_flux}(b). Since the relevant parameter is the minimum energy gap between states inside and outside the computational space during the implementation of an operation, leakage errors can occur even in the absence of $1/f$ noise. When $1/f$ noise is present, the minimum gaps are shifted from their ideal values when there is no noise, which in turn has an effect on  leakage.

We model a noisy preparation of a qubit in the state $\ket{+}$ as the ideal preparation followed by the noise superoperator
\begin{equation}
\label{eq:super}
{\cal N} \approx {\cal N}_{T_1} \circ {\cal N}_{1/f} \; ,
\end{equation}
\noindent where $\circ$ denotes composition. We model a noisy {\sc cphase} gate as the ideal gate followed by a superoperator as in Eq.~(\ref{eq:super}), where ${\cal N}_{1/f}$ is now supported on both qubits acted upon by the gate, and ${\cal N}_{T_1}$ is assumed to act independently on each of the two qubits. Finally, we model a noisy measurement of $\exp(i \theta \sigma_{\rm z})\sigma_{\rm x}$ as the ideal measurement {\em preceded} by a superoperator as in Eq.~(\ref{eq:super}).

The noise superoperator for each elementary operation can be expressed in terms of a discrete set of Kraus operators (at most $d^2$ where $d$ is the dimension of its support) \cite{preskill_chap7}. If the identity is one of the Kraus operators, the noise model corresponds to local stochastic noise; in this case, we may define the phase error rate as the probability of all non-identity Kraus operators which are diagonal matrices in the computational basis. The probability of all other non-identity Kraus operators then defines the rate for all other types of errors. For this noise model, we can perform a fault-tolerance analysis to determine upper bounds on the probabilities of logical errors as in \cite{Aliferis07d}.

However, as we will discuss below, the superoperators in our modeling do not have this property so that noise cannot be simply described in terms of probabilistic errors. Nonetheless, we may use an alternative definition of an error rate, and with this definition the analysis in \cite{Aliferis07d} remains essentially unchanged. The idea is to express $\mathcal{N}$ as the sum of an ideal and an erroneous part, ${\cal N}={\cal \hat I}+{\cal E}$, where ${\cal \hat I}$ is a trace-decreasing superoperator which is proportional to the identity superoperator \cite{AB:FTlong}. We may then define a generalized error rate or error {\em strength} in terms of the distance between ${\cal N}$ and ${\cal \hat I}$,
\begin{equation}
\varepsilon \equiv ||{\cal E}||_{\diamond} = ||{\cal N}-\mathcal{\hat I}||_{\diamond} \; ,
\end{equation}
\noindent where $||\cdot||_{\diamond}$ is the diamond norm \cite{KSV:computation}. If the superoperator ${\cal E}$ has an $n$-qubit input, $||{\cal E}||_{\diamond}=||{\cal I}_n \otimes {\cal E}||_1=  \max_{X\,:||X||_{\rm tr} =1} ||({\cal I}_n \otimes {\cal E})(X)||_{\rm tr}$, where ${\cal I}_n$ is the identity superoperator on $n$ qubits, and $||\cdot||_{\rm tr}$ is the standard trace norm, {\em i.e.}, $||A||_{\rm tr}={\rm Tr} \sqrt{A^{\dagger} A}$.

In the remainder of this Supplementary Material, we give the details about the error rates for our elementary operations; the results are summarized in the end in Table \ref{table:1}. To simplify our calculations, we have estimated the various norms by only varying among a few possible inputs. Even though we have not performed a rigorous maximization, we believe that the inputs we have chosen are close to the worst case.

\cleardoublepage
%------------------------------------------------------------------%
\vspace{0.3cm}
\noindent{\em $1/f$ noise in {\sc cphase} gates}

For the {\sc cphase} gate, we have obtained the combined superoperator ${\cal N}_{1/f} \circ {\cal CZ}$ by integrating over the Gaussian fluctuations in the applied fluxes and the pulse timing as described in the main text; here, ${\cal CZ}$ is the ideal {\sc cphase} superoperator (up to diagonal rotations on each qubit which are all moved before the measurements and will be discussed separately).
%i.e. the superoperator is a stochastic mixture of unitary gates.
From these numerical simulations, we can then extract the Kraus operators for ${\cal N}_{1/f}$. Each Kraus operator is supported on a 16-dimensional space $\mathcal{H}^{AB} = \mathcal{H}^A \otimes \mathcal{H}^B$, where $\mathcal{H}^A$ and $\mathcal{H}^B$ are 4-dimensional spaces corresponding to the two interacting Koch qubits A and B. For both qubits A and B,
\begin{equation}
\label{eq:tensor}
\mathcal{H}^Q = \mathcal{H}^Q_{\rm flux} \otimes \mathcal{H}^Q_{\rm trans} \; ,
\end{equation}
\noindent where $\mathcal{H}^Q_{\rm flux}$ is a 2-dimensional space spanned by the flux-qubit states $\{ |L\rangle, |R\rangle\}$ and $\mathcal{H}^Q_{\rm trans}$ is another 2-dimensional space spanned by the 0- and 1-photon states of the transmission line \footnote{Since transitions to states with more than one photon in a transmission line are negligible, in our numerical simulations we truncate the infinite-dimensional space of the transmission-line modes at the first excited state; see \cite{Brito07}.}. Since prior to and after the implementation of the {\sc cphase} gate information is stored in the transmission-line modes, the computational space corresponds to the tensor product of the two transmission-line spaces, $\mathcal{H}^{AB}_{\rm trans} = \mathcal{H}^A_{\rm trans} \otimes \mathcal{H}^B_{\rm trans}$; the action of the gate on the space $\mathcal{H}^{AB}_{\rm flux} = \mathcal{H}^A_{\rm flux} \otimes \mathcal{H}^B_{\rm flux}$ is ideally trivial, and any transfer of amplitude to this space corresponds to leakage.

We have found that, within the precision of our numerical analysis \footnote{We use 8 decimal digits of accuracy.}, only four Kraus operators carry significant weight while all the rest are negligible. These four Kraus operators $\{M_0, \dots, M_3 \}$ are of the form
\begin{equation}
M_k = I_k + M_{k,d} + M_{k, \neg \, d} + M_{k,l} \; .
\end{equation}
\noindent Here, $I_k$ is proportional to the identity operator on $\mathcal{H}^{AB}$, $M_{k,d}$ includes all terms that act as the identity on $\mathcal{H}^{AB}_{\rm flux}$ and as diagonal matrices in the computational basis on $\mathcal{H}^{AB}_{\rm trans}$ (giving rise to phase errors), $M_{k, \neg \, d}$ contains all remaining terms acting as the identity on $\mathcal{H}^{AB}_{\rm flux}$ (giving rise to other types of errors in the computational basis), and finally $M_{k,l}$ includes all terms that act non-trivially on $\mathcal{H}^{AB}_{\rm flux}$ (giving rise to leakage errors).

If we expand in the Pauli basis in $\mathcal{H}^{AB}$, we find $\{ I_0= .9981 \exp(i1.2743) I^A \otimes I^B$, $I_1=0,I_2=0,I_3=0 \}$,
\begin{equation}
\begin{array}{rcrrr}
%M_{0,d} & = & .9981 \; I^A\otimes I^B & &  \\
%    &   & \hspace{-.2cm}+(.44 - 1.44i) 10^{-4} \;  I^A \otimes \sigma_{\rm z}^B & + (3.64 + .06i)10^{-4} \; \sigma_{\rm z}^A \otimes I^B & - (4.56 +.13i)10^{-4} \; \sigma_{\rm z}^A \otimes \sigma_{\rm z}^B \; ,  \\
M_{0,d} & = & 1.5\times 10^{-4} \;  I^A \otimes \sigma_{\rm z}^B & + (1 + 3.5i)10^{-4} \; \sigma_{\rm z}^A \otimes I^B & - (1.2 + 4.4i)10^{-4} \; \sigma_{\rm z}^A \otimes \sigma_{\rm z}^B \; ,  \\ % I = .30 + .96i
M_{1,d} & = & 5.2\times 10^{-2} \; I^A \otimes \sigma_{\rm z}^B & + 9 \times 10^{-3} \; \sigma_{\rm z}^A\otimes I^B & - 7\times 10^{-3} \; \sigma_{\rm z}^A \otimes \sigma_{\rm z}^B \; , \\
M_{2,d} & = & 1.8\times 10^{-3} \; I^A \otimes \sigma_{\rm z}^B & + 1\times 10^{-2} \; \sigma_{\rm z}^A \otimes I^B & + 4.6\times 10^{-4} \;  \sigma_{\rm z}^A \otimes \sigma_{\rm z}^B \; , \\
M_{3,d} & = & 1\times 10^{-4} \; I^A\otimes \sigma_{\rm z}^B & & +(7.4-i) 10^{-4} \; \sigma_{\rm z}^A\otimes \sigma_{\rm z}^B \; ,
\end{array}
\end{equation}
\noindent where $I^A = I^B = I \otimes I$ and $\sigma_{\rm z}^A = \sigma_{\rm z}^B = I\otimes \sigma_{\rm z}$ according to the tensor-product structure in Eq.~(\ref{eq:tensor}). For brevity, we will omit the expressions for $\{M_{k, \neg \, d}\}$ and $\{M_{k,l}\}$.

We may write $\mathcal{N}_{1/f} = {\cal \hat I} + \mathcal{E}_l + \mathcal{E}_{\neg d} + \mathcal{E}_d$, where ${\cal \hat I} (X)= \sum_{k=0}^3 I_k X I_k^\dagger$
%
%\begin{equation}
% {\cal \hat I} (X)= \sum_{k=0}^3 I_k X I_k^\dagger \;
%\end{equation}
%
\noindent is trace-decreasing and proportional to the identity superoperator, $\mathcal{E}_l$ contains all terms with at least one insertion of a leakage error $\{M_{k,l} \}$, {\em i.e.},
\begin{equation}
{\cal E}_l (X)= {\cal N}_{1/f}(X) - \sum_{k=0}^3 (I_k + M_{k,d} + M_{k, \neg d}) X (I_k + M_{k,d} + M_{k, \neg d})^\dagger \; ,
\end{equation}
\noindent ${\cal E}_{\neg d}$ contains all remaining terms with at least one insertion of a non-dephasing error $\{M_{k,\neg d} \}$, {\em i.e.},
\begin{equation}
{\cal E}_{\neg d} (X) = ({\cal N}_{1/f} - {\cal E}_l)(X) - \sum_{k=0}^3 (I_k + M_{k,d}) X (I_k + M_{k,d})^\dagger \; ,
\end{equation}
\noindent and $\mathcal{E}_d$ contains all remaining terms with at least one insertion of a phase error $\{M_{k,d} \}$, {\em i.e.},
\begin{equation}
{\cal E}_{d} (X) = ({\cal N}_{1/f} - {\cal E}_l- {\cal E}_{\neg d})(X) - \sum_{k=0}^3 I_k X I_k^\dagger \; .
\end{equation}

We define the rate of leakage errors as the norm of $\mathcal{E}_l$; by taking as the worst-case input the state $\ket{\tilde 1}\ket{\tilde 1}$ ({\em cf.} Fig.~\ref{fig:band_flux}(b)), we find $||\mathcal{E}_{l}||_{\diamond} \approx 3.5\times 10^{-6}$. Similarly, we define the rate of non-dephasing errors as the norm of $\mathcal{E}_{\neg d}$; we find $||\mathcal{E}_{\neg d}||_{\diamond} = {\it O}(10^{-7})$, which can be neglected since it is much smaller than the contribution due to relaxation to be discussed below.

%By a similar procedure as for phase errors, we have in fact obtained separate leakage error rates for qubits A and B; for qubit A we find a rate of approximately $1.6\times 10^{-8}$, while for qubit B we find a significantly larger rate of approximately $3.5\times 10^{-6}$. %\marginpar{\footnotesize input state \\ for 2nd order \\leakage}

We finally define the rate for phase errors as the norm of ${\cal E}_d$. By varying among several possible inputs, we obtained the largest value of $||{\cal E}_{d} ||_{\diamond} \approx 4.73 \times 10^{-3}$ for the Bell state
\begin{equation}
\ket{\Phi_0} = \frac{1} {\sqrt{2}} (\ket{\tilde 0}\ket{\tilde 0}+\ket{\tilde 1}\ket{\tilde 1}) \; .
\end{equation}
\noindent Here, $|\tilde{0} \rangle \equiv |S\rangle |0_p\rangle$ and $|\tilde{1}\rangle \equiv |S\rangle |1_p\rangle$, where $|S\rangle = {1\over \sqrt{2}}(|L\rangle + |R\rangle)$ is the symmetric state in $\mathcal{H}^Q_{\rm flux}$ and $|n_p\rangle$ is the $n$-photon state in $\mathcal{H}^Q_{\rm trans}$ with $Q$ either A or B; see Figs.~\ref{fig:2} and \ref{fig:band_flux}.

We may also estimate the rates for phase errors on qubit A and qubit B separately. For qubit A, we modify our expansion by writing $\mathcal{N}_{1/f} = {\cal \hat I}^A + {\cal E}_{l} + {\cal E}_{\neg d} +  {\cal E}_d^A$, where
\begin{equation}
{\cal \hat I}^A(X) = \sum_{k=0}^3 (I_k + M_{k,d}^B) X (I_k + M_{k,d}^B)^{\dagger} \; ,
\end{equation}
\noindent and $M_{k,d}^B$ includes those terms in $M_{k,d}$ which are proportional to $I^A\otimes \sigma_{\rm z}^B$ and so act trivially on qubit A. Then, ${\cal E}_d^A$ captures all terms that apply nontrivial phase noise to qubit A. By using the same Bell state as input, we find that phase errors on qubit A have a rate $||\mathcal{E}^A_{d}||_{\diamond} \approx 1.96 \times 10^{-3}$. If we perform a similar analysis for qubit B instead, we find $||\mathcal{E}^B_{d}||_{\diamond} \approx 4.6 \times 10^{-3}$. This shows that the effect of $1/f$ noise on qubit $B$ is stronger than on qubit $A$, which is physically expected since qubit A is unparked much less deeply than qubit B during the implementation of a {\sc cphase} gate; see Fig.~\ref{fig:band_flux}.

%------------------------------------------------------------------%
\vspace{0.3cm}
\noindent{\em $1/f$ noise in preparation}

For the preparation of the state $|+\rangle$, we have obtained the density matrix $\rho_{\ket{+}} = {\cal N}_{1/f}(|+\rangle \langle +|)$ by performing a similar integration over the fluctuating fields as for the {\sc cphase} gate. Depending on the species of qubit, $\rho_{\ket{+}}$ is supported on $\mathcal{H}^Q$ for $Q$ either A or B, and it is of the form
\begin{equation}
\rho_{\ket{+}} = \eta_{\ket{+},d} + \eta_{\ket{+},l} \; ;
\end{equation}
\noindent here, $\eta_{\ket{+},d}$ is an operator supported on $|S\rangle \langle S| \otimes \mathcal{H}^Q_{\rm trans}$ , and $\eta_{\ket{+},l}$ includes all remaining terms of $\rho_{\ket{+}}$. Since after the preparation the information is stored in the transmission-line modes, the ideal state is $|\tilde +\rangle \langle \tilde +| = |S\rangle \langle S| \otimes |+_p\rangle \langle +_p|$ from which we can obtain $\eta_{\ket{+},d}$ by acting with phase errors alone; on the other hand, for all terms in $\eta_{\ket{+},l}$ the state of the flux qubit is orthogonal to $|S\rangle$ so that leakage has occurred.

We define the rate of leakage errors as $\varepsilon_l = ||\eta_{\ket{+},l}||_{\rm tr}$; for qubit A we find $\varepsilon_l \approx 3.77\times 10^{-7}$, while for qubit B we find $\varepsilon_l \approx 1.5 \times 10^{-5}$ (our pulses are not highly optimized to avoid leakage, and we believe the leakage error rate for qubit B can be improved if necessary). For phase errors, we define the rate as $\varepsilon = ||\eta_{\ket{+},d} - c|\tilde +\rangle \langle \tilde +|||_{\rm tr}$ where we are allowed to optimize over the choice of $0\leq c\leq 1$; for both qubits A and B, we find $\varepsilon \approx 2.75\times 10^{-3}$.

%------------------------------------------------------------------%
\vspace{0.3cm}
\noindent{\em $1/f$ noise in measurement}

A measurement of $\exp(i\theta\sigma_{\rm z})\sigma_{\rm x}$ is implemented by applying the diagonal rotation $\exp(-i \theta \sigma_{\rm z})$ on $\mathcal{H}^Q_{\rm trans}$, followed by a non-adiabatic pulse mapping $\ket{\tilde +}$ and $\ket{\tilde -}$ (where information is stored in the transmission-line modes) to $|0\rangle \equiv |L\rangle |0_p\rangle$ and $|1\rangle\equiv |R\rangle |0_p\rangle$ respectively (where information is stored in the flux-qubit states), followed by a projection along the basis $\{|L\rangle, |R\rangle\}$. %Since any leakage during this process will transfer amplitude to excited states of the transmission line in the final state before the projection, the measurement cannot distinguish whether leakage has occurred.

Diagonal rotations can be executed with pulses of short duration so that errors are very weak compared to other operations and can be neglected. We also assume that errors during the final projection can be neglected. The states $|L\rangle$ and $|R\rangle$ can be distinguished very accurately by setting $\Phi_C$ very small ($\sim 1.4 \Phi_0$) so that there is a large potential barrier between them resulting in very high $T_1$ in the $\{|L\rangle, |R\rangle\}$ basis ({\em  cf}. Fig.~\ref{fig:2}); since the two states correspond to distinct circulating-current orientations, they induce different magnetic signals which can be detected by using the SQUIDs ({\em  cf}. Fig.~\ref{fig:1}).

To obtain the error rate for the remaining measurement process, we follow the evolution of the basis states $\ket{\tilde +}$ and $\ket{\tilde -}$ by performing a numerical simulation similar to the cases already discussed. If the initial state is $\ket{\tilde +}$, we calculate the probability that the final state before the projection is orthogonal to $|0\rangle$, in which case we assume an error in the measurement outcome always occurs; and similarly for $\ket{\tilde -}$, we calculate the probability that the final state is orthogonal to $|1\rangle$. We define this probability as the error rate for the measurement; for both qubits A and B, we find  $\varepsilon \approx 1.83\times 10^{-3}$.

%------------------------------------------------------------------%
\vspace{0.3cm}
\noindent{\em Relaxation}

We model relaxation noise by the amplitude-damping superoperator $\mathcal{N}_{T_1}$ acting independently on each qubit; with $Q$ either A or B, the two Kraus operators are
\begin{equation}
%A_0=\left(\ba{cc} 1 & 0 \\ 0 & \sqrt{1-\gamma} \ea\right) \; , \; \;
M_0 = {1 + \sqrt{1-\gamma} \over 2} \; I^Q + {1-\sqrt{1-\gamma}\over 2} \; \sigma_{\rm z}^Q \; , \;\;
%A_1=\left(\ba{cc} 0 & \sqrt{\gamma} \\ 0 & 0 \ea\right) \; ,
M_1 = {\sqrt{\gamma}\over 2} \sigma_{\rm x}^Q (1 - \sigma_{\rm z}^Q) \; ,
\end{equation}
\noindent where we have already defined $I^Q$ and $\sigma_{\rm z}^Q$, and $\sigma_{\rm x}^Q = I \otimes \sigma_{\rm x}$ according to the tensor-product structure in Eq.~(\ref{eq:tensor}).

Since only $M_1$ is non-diagonal in the computational basis, we define the rate for non-dephasing errors as $
||\mathcal{M}_1 ||_\diamond = \gamma$, where $\mathcal{M}_1 (X) = M_1 X M_1^\dagger$. We also define the rate for phase errors due to the operator $M_0$ as $||\mathcal{M}_0 - {\cal \hat I}||_\diamond$, where $\mathcal{M}_0 (X) = M_0 X M_0^\dagger$ and ${\cal \hat I}(X) = cX$ with $0\leq c\leq 1$. If we take $c= \left(1 + \sqrt{1-\gamma}\right)^2/4$, we find $||\mathcal{M}_0 - {\cal \hat I}||_\diamond \approx \gamma/2$.

For a {\sc cphase} gate, we use the worst-case estimate $T_1 = 10m{\rm sec}$, and we assume $T_1$ can be treated as approximately constant during the execution of the gate; then, $\gamma=t/T_1 = 3.5\times 10^{-6}$ where $t = 35n{\rm sec}$ is the duration of the gate. For diagonal rotations which are executed with pulses of duration $t\leq 5n{\rm sec}$, $\gamma$ is very small and can be neglected. Finally, for preparation and measurement, $T_1$ changes as a function of the control flux and we calculate $\gamma = \int_{0}^t ds \frac{s}{T_1(s)} = 3.5 \times 10^{-7}$.

%------------------------------------------------------------------%

\begin{table}[t]
\begin{center}
\begin{tabular}{r|c|cc}
                  &  & {\footnotesize Qubit A ($\omega_T=2 \pi \times 3.1$ GHz)} & {\footnotesize Qubit B ($\omega_T=2\pi \times 3/4 \times 3.1$ GHz)}\\
\hline \hline
{\sc cphase} &
$\varepsilon$                    & $1.96 \times 10^{-3}$             & $4.6 \times 10^{-3}$ \\
& $\varepsilon'$                 & $3.5\times 10^{-6}$               & $3.5\times 10^{-6}$ \\
& $\varepsilon_{l}$              & \put(105,0){$3.5 \times 10^{-6}$}  \\
\hline \hline
$\ket{+}\;{\rm prep.}$ &
$\varepsilon$                    & $2.75 \times 10^{-3}$             & $2.75 \times 10^{-3}$ \\
& $\varepsilon'$                 & $3.5\times 10^{-7}$               & $3.5\times 10^{-7}$ \\
& $\varepsilon_{l}$              & $3.77 \times 10^{-7}$             & $1.5\times 10^{-5}$  \\
%\hline \hline
%$\exp(i\theta\sigma_{\rm z})$:   &
%$\varepsilon$                    & \put(112,0){$\approx 0$}          & \\
%& $\varepsilon'$                 & \put(106,0){$5 \times 10^{-8}$}   & \\
%& $\varepsilon_{l} $             & \put(112,0){$\approx 0$}          & \\
\hline \hline
$\exp(i\theta\sigma_{\rm z}) \sigma_{\rm x}\;{\rm meas.}$   &
$\varepsilon$                    & $1.83\times 10^{-3}$              & $1.83\times 10^{-3}$
\end{tabular}
\end{center}
\caption{\label{table:1} Error-rate estimates for our elementary operations. For preparations and {\sc cphase} gates, $\varepsilon$ is the rate for phase errors, $\varepsilon'$ is the rate for all other types of errors, and $\varepsilon_l$ is the rate for leakage errors. For measurements, the rate $\varepsilon$ includes errors from all sources. }
\end{table}

%------------------------------------------------------------------%
%\vspace{0.3cm}
%\noindent{\em Fault-tolerance analysis}

%We can expand the superoperator corresponding to a rectangle using the sum of an ideal operator and an error operator for each location. For the superoperator ${\cal S}_{T_1}$ we assume that any single ${\cal E}_{T_1}$ is bad since the encoding is not meant to correct against these errors. Similarly, any single leakage error is considered bad. Hence the logical failure probability of an encoded CNOT gates is $\epsilon_{\rm bad} \leq \sum_{\mbox{fault-paths }F} ||{\cal S}_F||$.

%We will use properties of the norm $||{\cal S} \circ {\cal T}|| \leq ||{\cal S}|| ||{\cal T}||$ and for a trace-decreasing superoperator $||{\cal S}|| \leq 1$. Each location is expanded into a sum of different error operators corresponding to different events. In this way we obtain a tree of event and one needs to sum over all the bad branches of this tree. If a branch leads to a bad event (say, a single leakage error), we terminate the branching and append superoperators for the remaining locations \cite{AB:FTlong}.

%-----------------------------------------------------------------------------------------------------------%
\bibliographystyle{unsrt}

\end{document}